\documentclass[a4paper,11pt]{scrartcl}

\usepackage{graphicx}
\usepackage{amsmath,amssymb,amsthm}
\usepackage[left=18mm,top=18mm]{geometry}
\usepackage{alltt}
\usepackage{fancyvrb}
\usepackage{datetime2}
\usepackage{authblk}

\newcommand{\refsec}[1]{Section~\ref{#1}}
\newcommand{\reftab}[1]{Table~\ref{#1}}

\newcommand{\nbcode}[1]{\mbox{\texttt{#1}}}
\newcommand{\notSomething}[1]{\overline{#1}}
\newcommand{\notX}{\notSomething{x}}
\newcommand{\notXaddmath}{\ensuremath{\notX{}}}

\setcapindent{1em} 

\title{Low latency data-flow graphs for simultaneous modular inversion
of many inputs}

\author{Tamás Visegrády}
\affil{Ripple Labs Research \\
       \texttt{tvisegrady@ripple.com}}
\date{\today}

\begin{document}

\maketitle

\section*{Abstract}

Montgomery's trick accelerates simultaneous modular inversion of
$N$ inputs by amortizing a single shared inversion, but auxiliary
multiplications for complement products are typically scheduled in a
linear, serial form. We construct a maximally parallelizable data-flow
graph (DFG) that computes all $\overline{x}$ complement~products by
scheduling auxiliary multiplications into idle multiplier slots during
accumulation of the product of all inputs, and that of the shared
inversion. This scheduling ensures the post-inversion phase adds
exactly one multiplication layer of latency regardless of $N$, yielding
a critical path latency of $\lceil \log_2 N \rceil$ multiply layers,
one inversion, and one final parallel multiply layer.

Measurements on an AMD~Ryzen AI PRO 350 using \nbcode{libsecp256k1}
primitives yield $1.27\times$ single-inversion latency for $N=8$
and $1.34\times$ for $N=16$, with peak multiplier load ($>N/2$
simultaneous multiplications) lasting only 80~ns out of roughly 1500--1600~ns
total---considerably less than replicating $N$ inversions would require.
The construction suits latency-sensitive batch workloads such as those
related to MPC presignature generation; it supports constant-time
execution when the underlying primitives are constant-time.


\section{Overview}

Simultaneous modular inversions using the same modulus may
be accelerated using ``Montgomery's trick,'' adding modular
multiplications to amortize the high cost of a single shared~inversion
\cite[4.1]{pairaffine2010}. This batching technique, extrapolated
from a few inputs, is typically implemented with linear sequences
of multiplications \cite{rust-multiinvert}. Significant latency
improvements have been demonstrated for single-input inversion
\cite[1.2]{byanginvert2026}; we consider latency reduction specific to
the multi-input case.

For inputs $a$ and $b$, Montgomery's~trick calculates $Q=a \cdot b$,
$1/Q=1/(a \cdot{} b),$ $1/a=b \cdot 1/Q,$ and $1/b=a \cdot 1/Q,$ saving
an inversion while adding three multiplications. Straightforward
extrapolation to a higher number of inputs tends to be implemented in an
inherently serial form.

For many simultaneous~inversions, one may construct an equivalent,
parallelizable data-flow~graph (DFG) which decreases overall latency
at the cost of adding many simultaneous multiplications. We use
\notXaddmath{}---``\nbcode{notX}'' in figures---to denote the
product of all inputs except~$x,$ so $1/x=\overline{x} \cdot 1/Q$ ($Q$
is the full product). \emph{We schedule auxiliary multiplications
for $\overline{x}$ in idle periods for multipliers,} such that
post-inversion steps add only one multiplication's latency, regardless
of the number of outputs---if sufficient multipliers are available.

Simultaneous inversion necessarily computes $Q$ and all its partial
products on the critical path. With two-input modular~multipliers as
building blocks, the critical~path includes $\lceil \log_2 N \rceil$
layers of multiplications \cite{montginvtree} for $N$ inputs.
The constant single-multiplication latency of final $1/x=\overline{x}
\cdot 1/Q$ products is an improvement over straightforward serial or
even tree-structured reconstruction \cite{montginvtree}, made
possible by computing products while inversion is running.


\emph{Parallel computation of \notXaddmath{} terms is suitable for
cases where bursts of peak load may be tolerated, and low latency
is at a premium.} Certain forms of batched presignature~generation
for MPC~protocols \cite[6]{ku23mpc}, or inverting $k$~nonces for
batch-generated ECDSA~signatures are potential use cases where one
may consider multi-input inversion. While we may use up to $N$
multipliers simultaneously for minimal latency, since we parallelize
only multiplications---but not the longer-running inversion---we
generate shorter periods of peak load than replicating $N$ inversions
would do.

\subsection{Scheduling multiplications not on the critical path}

\begin{figure}
\begin{center}
\resizebox{\textwidth}{!}{\includegraphics{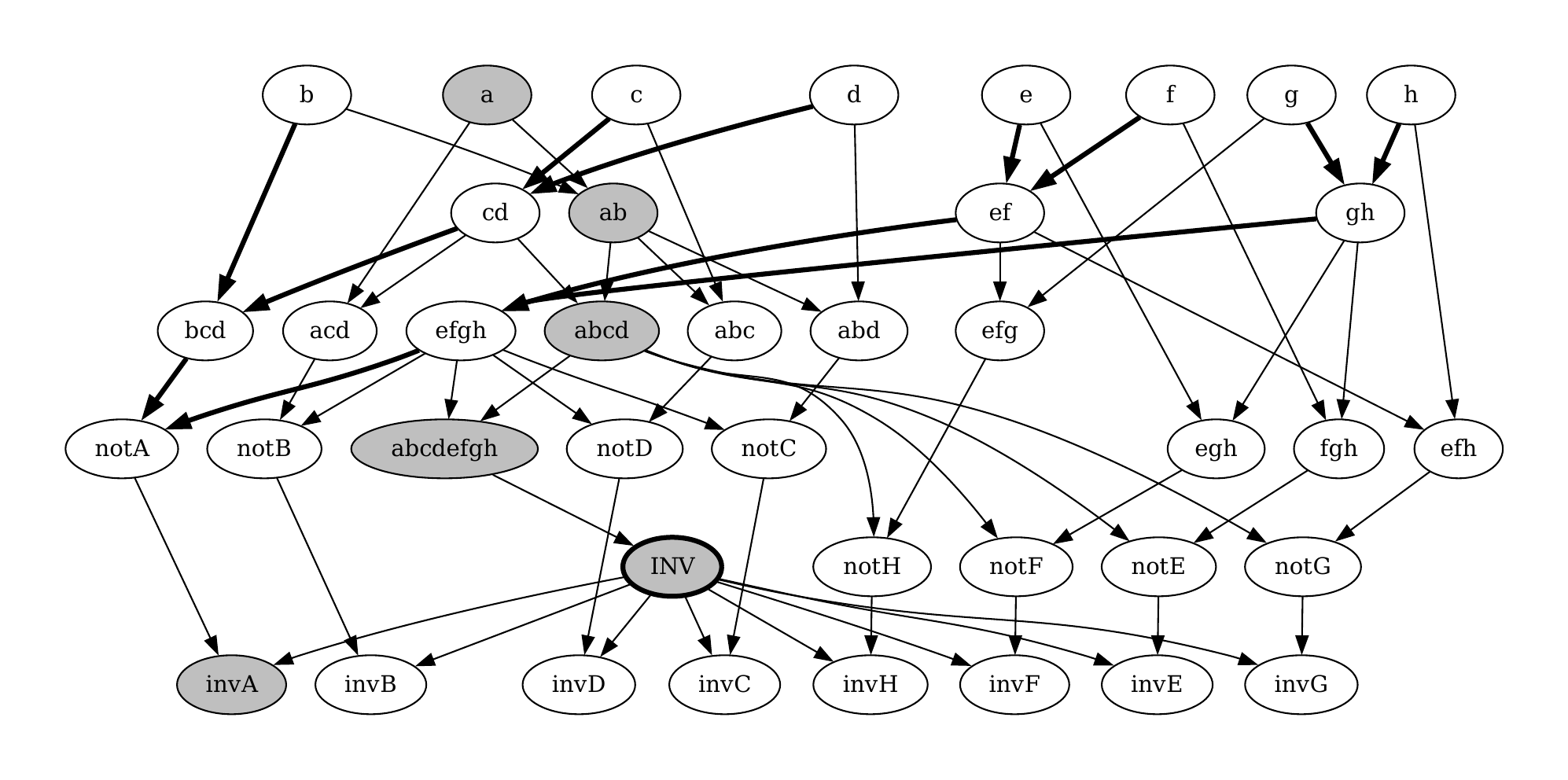}}
\caption{Parallel data-flow graph for 8 simultaneous inversions}
\label{fig:dfg8}
\end{center}
\end{figure}

Unlike $Q,$ not all multiplications of $\overline{x}$ are on the
critical path. There are several reasons the schedule of products
included in $\overline{x}$ terms is less restricted, given a fixed number
of multipliers:

\begin{enumerate}
\item the number of multipliers used by dependencies of $Q$ is halved in
each layer of the initial multiplication tree.

In the Fig.~\ref{fig:dfg8} example, the first $Q$ layer computes four
products: $a b,$ $c d$, $e f,$ and $g h$. The next layer may not utilize
more than two multipliers: it only calculates $a b c d$ and $e f
g h.$
\label{i:decreasing}

\item \notXaddmath{} combines products from different layers of the
initial multiplication tree which supplies $Q.$ Accumulation of those
terms may be started as soon as some of the products are available.

As an example, $\notSomething{a}$ with $N=8$ may already multiply $b$
with $cd$ in the second layer while $efgh$ is being computed on the
critical path; we may immediately produce $\notSomething{a}$ in the
next layer (Fig.~\ref{fig:dfg8}). With the increasing availability of
multipliers due to (\ref{i:decreasing}) and the $Q$ tree generating many
small partial products by this point, there is considerable
flexibility starting \notXaddmath{} products' accumulation without
impacting the critical path.

\item since \notXaddmath{} products are only needed when they are to
be multiplied with $1/Q,$ we may schedule their multiplication chains
during the calculation of the inverse.
\end{enumerate}

\begin{figure}
\begin{center}
\resizebox{\textwidth}{!}{\includegraphics{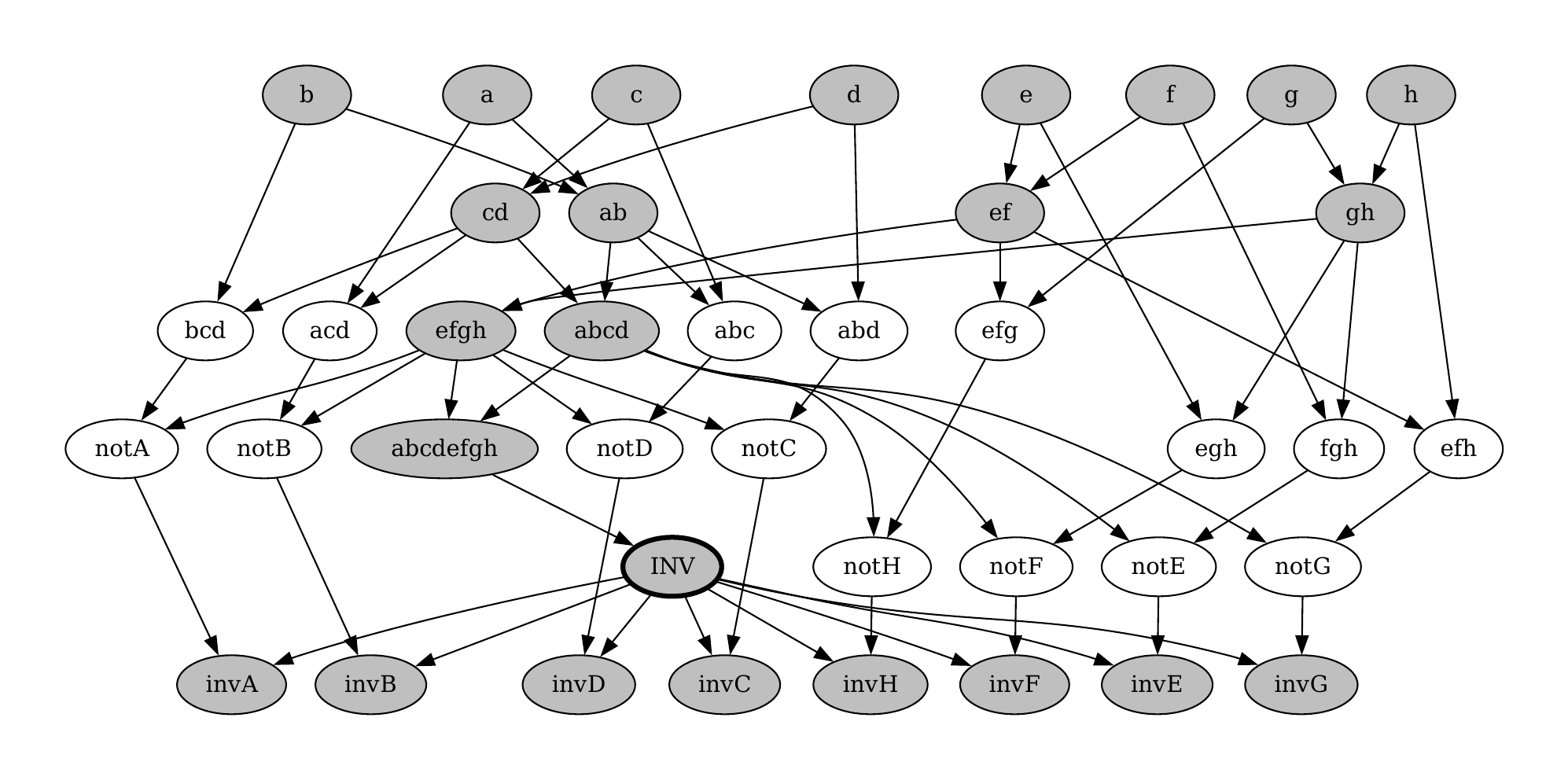}}
\caption{Estimated critical path for 8 simultaneous inversions}
\label{fig:dfg8critpath}
\end{center}
\end{figure}

\begin{figure}
\begin{center}
\resizebox{\textwidth}{!}{\includegraphics{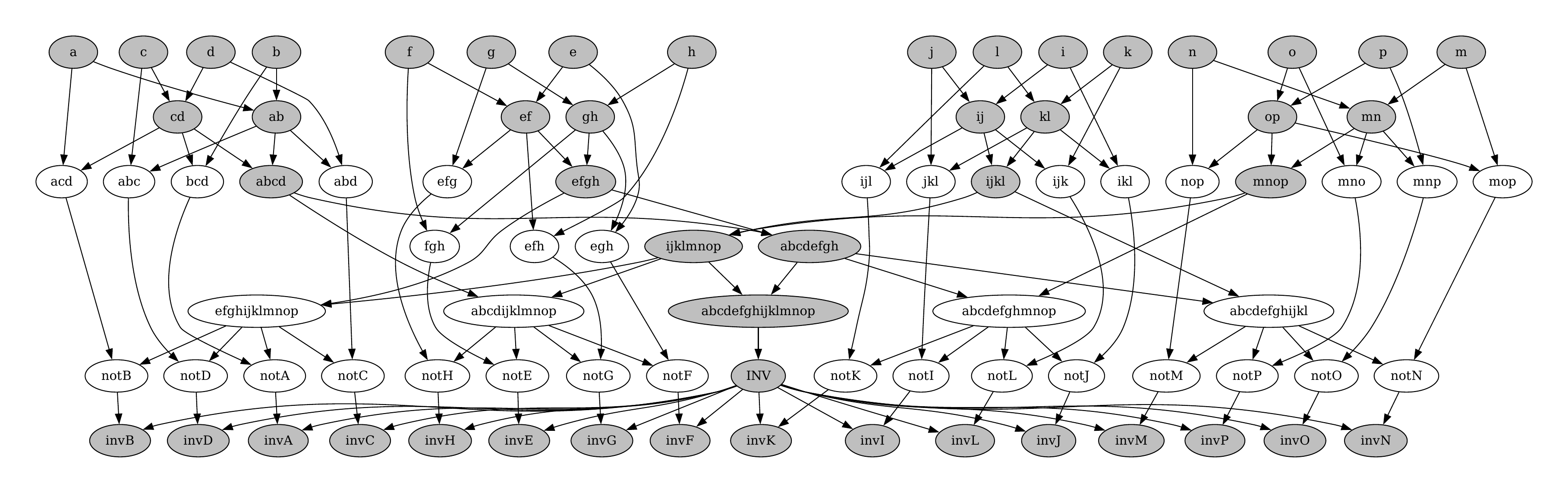}}
\caption{Estimated critical path for 16 simultaneous inversions}
\label{fig:dfg16critpath}
\end{center}
\end{figure}

\section{Examples for $N=8$ and $N=16$ inputs}

Fig.~\ref{fig:dfg8} shows the terms of simultaneous
inversion of $N=8$ inputs $a \ldots h$:
\begin{align*}
Q            &=a \cdot b \cdot c \cdot d \cdot e \cdot f \cdot g \cdot h \\
\overline{a} &=b \cdot c \cdot d \cdot e \cdot f \cdot g \cdot h         \\
1/a          &=\overline{a} \cdot 1/Q
\end{align*}

Operations highlighted in gray illustrate the $1/a$---``\nbcode{invA}''
in figures---data-flow:
\begin{itemize}
\item construct the full~product $Q$ in $\lceil \log_2 N \rceil$ layers of
      multiplications. Up to $\lceil N/2^i \rceil$ simultaneous
      multiplications may be scheduled in layer~$i.$
\item compute $1/Q$ with a single modular~inversion.
\item generate $\notSomething{a}$. \emph{Not necessarily on the critical path.}
\item generate $1/a=\notSomething{a} \cdot 1/Q$ etc. in a single layer
      of final multiplications. These $N$ final multiplications are
      independent, and may simultaneously utilize up to $N$ multipliers.

      If only $M<N$ multiplications may be scheduled simultaneously,
      post-inversion multiplications complete in $\lceil N / M \rceil$
      layers.
\end{itemize}

Dependencies of $Q$ are power-of-two product terms; all are on the
critical path. Additional products are included in \notXaddmath{},
combined with some of the $Q$ terms---such as $b \cdot \left( c \cdot
d \right)$ for $1/a;$ those dependencies are shown with wider arrows for $\overline{a}$
in Fig.~\ref{fig:dfg8}. These additional products are typically not on
the critical path: they only need to finish before they are used by the
final multiplication which produces $1/x$---without increasing overall
latency.

Fig.~\ref{fig:dfg8critpath} shows critical-path operations in gray, and
those which may be scheduled somewhat freely in white.
Horizontal layers of multiplications in both Fig.~\ref{fig:dfg8} and
Fig.~\ref{fig:dfg8critpath} show how a parallelized environment with
$M=8$ multipliers could be sufficient: we schedule layers such that the
number of simultaneous multiplications is never over~$8.$

Fig.~\ref{fig:dfg16critpath} shows the critical and non-critical-path
operations of a comparable data-flow graph for $N=16$ inputs. A sample
multiplier schedule in Table~\ref{tab:mul16-util} accommodates all data
dependencies of a 16-input simultaneous inversion if $M=8$ multipliers
may be simultaneously utilized. Partial products of \notXaddmath{} are
redundantly parenthesized to show which product from $Q$ they combine;
critical~path multiplications---excluding the final $\notX \cdot
1/Q$ ones---are underlined.

If the $1/Q$ inversion of Table~\ref{tab:mul16-util} terminates
in at least four times the latency of a single multiplication---a
reasonable assumption (see \refsec{s:performance})---generating all
$\notX$ products does not increase overall latency beyond that of the
critical~path.


{
\begin{center}
\begin{table}
\scriptsize
\begin{alltt}
multiplier#1        #2            #3              #4             #5             #6           #7               #8
 computes:

     \underline{ab}             \underline{cd}            \underline{ef}              \underline{gh}             \underline{ij}             \underline{kl}           \underline{mn}               \underline{op}

    \underline{abcd}           \underline{efgh}          \underline{ijkl}            \underline{mnop}           b(cd)          a(cd)        (ab)d            (ab)c

  \underline{abcdefgh}       \underline{ijklmnop}        f(gh)           e(gh)          (ef)h          (ef)g        j(kl)            i(kl)

 \underline{abcdefgh..}        (ij)l         (ij)k           n(op)          m(op)          (mn)p        (mn)o            [idle]
 \underline{..ijklmnop}

----------------  1/Q=1/(a b .. o p) inversion starts; critical-path multiplications for Q finished  -----------------

 b(cd)(efgh)    a(cd)(efgh)    (ab)d(efgh)    (ab)c(efgh)    (abcd)f(gh)    (abcd)e(gh)    (abcd)(ef)h    (abcd)(ef)g

 j(kl)(mnop)    i(kl)(mnop)    (ij)l(mnop)    (ij)k(mnop)    (ijkl)n(op)    (ijkl)m(op)    (ijkl)(mn)p    (ijkl)(mn)o

b(cd)(efgh)..  a(cd)(efgh)..  (ab)d(efgh)..  (ab)c(efgh)..  (abcd)f(gh)..  (abcd)e(gh)..  (abcd)(ef)h..  (abcd)(ef)g..
 ..(ijklmnop)   ..(ijklmnop)   ..(ijklmnop)   ..(ijklmnop)   ..(ijklmnop)   ..(ijklmnop)   ..(ijklmnop)   ..(ijklmnop)

---------------------------------------------  notA .. notH available  -----------------------------------------------

(abcdefgh)..   (abcdefgh)..   (abcdefgh)..   (abcdefgh)..   (abcdefgh)..   (abcdefgh)..   (abcdefgh)..   (abcdefgh)..
..j(kl)(mnop)  ..i(kl)(mnop)  ..(ij)l(mnop)  ..(ij)k(mnop)  ..(ijkl)n(op)  ..(ijkl)m(op)  ..(ijkl)(mn)p  ..(ijkl)(mn)o

--------------------------------------  notI .. notP (all notX) available  -------------------------------------------

                                           ..after 1/Q is available..

1/a=notA*1/Q   1/b=notB*1/Q   1/c=notC*1/Q   1/d=notD*1/Q   1/e=notE*1/Q   1/f=notF*1/Q   1/g=notG*1/Q   1/h=notH*1/Q

1/i=notI*1/Q   1/j=notJ*1/Q   1/k=notK*1/Q   1/l=notL*1/Q   1/m=notM*1/Q   1/n=notN*1/Q   1/o=notO*1/Q   1/p=notP*1/Q

\end{alltt}
\caption{Multiplier utilization/generated products' schedule for $N=16$ inputs
         $a \ldots p$ and $M=8$ multipliers}
\label{tab:mul16-util}
\end{table}
\end{center}
}

\subsection{Other number of inputs}

If $N$ is not a power of two, some of the first layer of multiplications
for $Q$ are not used; no other changes are necessary. These freed
multiplier slots in the first layer may not be conveniently reused for
$\notX$ products. The structure of multiplications reveals the number of
inputs, which we consider public information, and therefore not consider
a security problem.

The final step computing $1/x$ is designed for maximal parallelism;
there is no change needed if $N$ is not a power of two.

\section{Implementation notes}

\paragraph{Local parallelization}
\label{s:multithread}

Since the parallelized DFG consists almost entirely of---reasonably
fast---multiplications using the same modulus, we expect software
implementations to utilize only fine-grained parallelism. Four-way
local parallelism has been demonstrated for comparable elliptic~curve
primitives \cite[Fig.~4]{ecc4x2020}, and single-CPU dispatch
of up to 8~units is available in state-of-the-art processors
\cite[2.9.7]{zen5opt2408} \cite[2.7.3]{x64opt2401}.

State-of-the-art processors have enough registers to support
a comparable modest number of simultaneous multiplications
\cite[3.4.1.1]{x86devref2602}, so we expect to be able to specialize
to this level of fine-grained parallelism. A specialized code block
which interleaves multiplying a few pairs of unrelated inputs
provides higher throughput than a loop invoking the same number of
individual multiplications. Our DFGs may use such a batched-multiplier
code~block, and utilize processor-level multi-dispatch, without invoking
higher-level parallelization.

Hardware platforms, if they can utilize lots of multipliers, may expand
$\notX$ dependencies with considerable freedom. With as many
multipliers as inputs---to accommodate post-inversion multiplications
as a single layer---the entire DFG may be instantiated with minimal
latency.

\paragraph{Specialized reuse of common subexpressions}
\label{s:furtherspec}

We considered, but did not evaluate multi-multiplication code blocks
specialized to the case of reused inputs. From our Fig.~\ref{fig:dfg8}
example, simultaneous computation of $(a b) (c d),$ $b (c d)$ and $a (c
d)$ could reuse intermediate values inside a multi-multiplication block.
Products for $\notX$ conveniently include terms in such a way that
common factors' batches-of-multiplications may be exploited---or,
alternatively, dispatched to a specialized code~block for $N$
simultaneous multiplications with considerably less than $2N$ inputs.

\paragraph{Centralized exception handling}

For environments where the inverse always exists---such as MPC trusted
dealers who may ensure shards are in valid~range---a single, centralized
divide-by-zero check serves as a ``should-not-happen'' exception.
Since multiplications are always successful, a single check for
\texttt{invert(0)} covers all reasonable exceptional conditions. We
expect multipliers and inversion to deal with incomplete reduction
\cite[4.2]{scott-ec24} or other representation details in a consistent
way, and may incorporate additional checks around them if necessary.

With inputs from untrusted sources and a prime modulus, checking for
zeroes in the first layer of multiplications ensures that the
\texttt{invert(0)} is not reached.

\paragraph{Constant or variable-time execution}

If an implementation has access to constant-time inversion
\cite[1.1]{byanginvert2026} and constant-time multiplication, our
DFG---fixed for any given $N$ input count---executes in constant time,
and uses a fixed memory-access pattern.


If a software implementation passes multiplications to a CPU-local
thread pool, or similarly utilizes multiple cores---the modest amount of
data allows this---we would not expect constant time execution. However,
inherent variation would not be data-dependent even in this case.

\section{Performance comparisons}
\label{s:performance}

We used a multi-core CPU to measure full latency of inverting
$N=8$ and $N=16$ values, using fast and constant-time multiply and
inverse primitives of the industry-standard \nbcode{libsecp256k1}
library \cite{libsecp256k1}. Modular inverse uses a state-of-the
art ``\nbcode{hddivsteps}'' \cite{byinvert-bounds} instantiation of
Bernstein-Yang inversion \cite{byanginvert2019}---significantly faster
than using modular exponentiation to invert in our test environment.

We measured both individual multiplications, and a multithreaded
wrapper where each thread multiplies two pairs of inputs per
call---this parallelizable form is applicable to our approach,
see~\refsec{s:multithread}. Multiplications from \nbcode{libsecp256k1}
are inlined, without register pressure or call~overhead; those in
their own threads are inherently slower. Effective throughput with
a separate thread interleaving a pair of multiplications per call
is approximately 55\% lower than that of minimal-overhead inlined
single multiplications, as we show below. While multi-threaded
multiplication is not expected to achieve the performance of
minimal-overhead single~multiplications, the latter provides a useful
performance baseline. \reftab{tab:inv-latencies} shows timings with
dual-multiplication threads, so the number of multiplication invocations
is half of multiplications used.

We tested performance with commit \nbcode{5698e66c} using a custom test
program directly calling primitives, and then separate dual-multiplying
threads, on an idle system with 100+ million force-serialized calls.
Median and mean were essentially identical for each multi-million
iteration loop. We compiled with gcc~15.3.0 from OpenSUSE Tumbleweed,
on an AMD Ryzen AI PRO~350 (Zen~5), with \nbcode{libsecp256k1}
default performance-build settings. This CPU has enough cores to run
post-inversion multiplications in parallel even for $N=16$, therefore
the final multiplications completed simultaneously---unlike the example
in \reftab{tab:mul16-util} which used fewer multipliers to illustrate a
schedule with lower peak~load.

Median latencies of inlined single multiplications, two-pair
multiplication in separate threads, and inversions were 25.8,
80.0 and 1186.3~ns, respectively. The ratio of latencies is 14.8
between inversion and two-pair multiplication in a different
thread, which allows a relaxed schedule for $\notX$ products.
Table~\ref{tab:inv-latencies} shows latency for our parallelized cases,
compared to single inversions.

For the $N$ values we tested, we were able to schedule auxiliary
multiplications for~$\notX$ with very low parallelism due to the
high inversion/multiplication latency ratio; only post-inversion
multiplications used more than $N/2$ parallel threads.

\begin{table}
\[
\begin{array}{|l||c||c|c|}
\hline
\text{number of inputs }(N)         &   1    &   8    &   16   \\ \hline\hline
\text{multiply layers computing }Q  &   0    &   3    &    4   \\ \hline
\text{inversion}                    &   1    &   1    &    1   \\ \hline
\text{post-invert multiply layers}  &   0    &   1    &    1   \\ \hline\hline
\text{full latency (ns)}            & 1186.3 & 1506.3 & 1586.3 \\ \hline
\text{relative latency}             &  1.0   &  1.27  &  1.34  \\ \hline\hline
\text{duration with load}>N/2\text{ (ns)}
                                    &  N/A   &  80.0  &  80.0  \\ \hline
\end{array}
\]
\caption{Inversion latencies with parallel, multithreaded multiplications}
\label{tab:inv-latencies}
\end{table}


\section{Conclusion}

We presented a parallelizable data-flow graph for simultaneous modular
inversion of $N$ inputs that extends Montgomery's trick by scheduling
auxiliary products in idle multiplier slots during the accumulation
of the product of all inputs, and the shared inversion. This ensures
the post-inversion phase adds exactly one multiplication layer of
latency---if sufficient multipliers are available. With $N$ multipliers
available in the final phase, the critical path consists of $\lceil
\log_2 N \rceil$ multiply layers, one inversion, and one final parallel
multiply~layer.

Measurements on an AMD Ryzen AI PRO~350 CPU using \nbcode{libsecp256k1}
show $1.27\times$ single-inversion latency for $N=8$ and $1.34\times$
for $N=16,$ with peak load ($>N/2$ multiplier threads
simultaneously active) lasting only 80~ns out of roughly 1500--1600~ns
total---considerably less than replicating $N$ inversions would require.
The construction suits latency-sensitive batch workloads such as MPC
presignature generation or ECDSA nonce inversion, and it is compatible
with constant-time execution when the underlying primitives are.

\bibliographystyle{alpha}
\bibliography{main}

\end{document}